\documentclass[preprintnumbers,twocolumn,prl,aps,showpacs,10pt,superscriptaddress,floatfix]{revtex4-1}

\usepackage[pdftex]{graphicx} 
\usepackage{epstopdf}
\usepackage{verbatim}
\usepackage{color}
\usepackage{subfigure}
\usepackage{amsmath, amsthm, amssymb}
\usepackage{hyperref}

\newcommand{\be}{\begin{equation}}
\newcommand{\ee}{\end{equation}}

\begin{document}
\title{Coulomb liquid phases of bosonic cluster Mott insulators on a pyrochlore lattice}
\author{Jian-Ping Lv}
\affiliation{Department of Physics, Anhui Normal University, Wuhu 241000, China}
\author{Gang Chen}
\email{chggst@gmail.com}
\affiliation{Collaborative Innovation Center of Advanced Microstructures and Department of Physics and Center for Field Theory and Particle Physics, Fudan University, Shanghai 200433, China}
\affiliation{Department of Physics, University of Toronto, Toronto, Ontario M5S1A7, Canada}
\author{Youjin Deng}
\affiliation{Hefei National Laboratory for Physical Sciences at Microscale and Department of Modern Physics,University of Science and Technology of China, Hefei 230027, China}
\author{Zi Yang Meng}
\email{zymeng@iphy.ac.cn}
\affiliation{Beijing National Laboratory for Condensed Matter Physics, and Institute of Physics, Chinese Academy of Sciences, Beijing 100190, China}

\date{\today}

\begin{abstract}
Employing large-scale quantum Monte Carlo simulations, we reveal the full phase diagram of the extended Hubbard model of hard-core bosons on the pyrochlore lattice with partial fillings.  When the inter-site repulsion is dominant, the system is in a cluster Mott insulator phase with integer number of bosons localized inside the tetrahedral units of the pyrochlore lattice. We show that the full phase diagram contains three cluster Mott insulator phases with $1/4$, $1/2$ and $3/4$ boson fillings, respectively. We further demonstrate that all three cluster Mott insulators are Coulomb liquid phases and its low-energy property is described by the emergent compact $U(1)$ quantum electrodynamics (QED). Besides measuring the specific heat and entropy of the cluster Mott insulators, we investigate the correlation function of the emergent electric field and verify it is consistent with the compact $U(1)$ quantum electrodyamics description. Our results shed lights on magnetic properties of various pyrochlore systems as well as the charge physics of the cluster magnets. 
\end{abstract}

  \maketitle

\paragraph*{Introduction.}
Cluster Mott insulators (CMIs) and related materials are new class of physical systems that provides a new ground to look for emergent and exotic phenomena~\cite{GangChen2014a,*GangChen2014b,*GangChen2015,Sheckelton2012,Abd-Elmeguid2004}. Unlike in the conventional Mott insulators, the electrons in CMIs are localized inside certain cluster units rather than on the lattice sites. It is the inter-site electron interaction that leads to the unconventional electron localization in CMIs and causes the electron motion inside neighboring clusters to be correlated. This correlated electron motion allows the electron charges to fluctuate quantum mechanically in the positional space at energies below the Mott gap. Such a sub-gap charge fluctuation is one of the key features of CMIs and is responsible for various emergent phenomena~\cite{GangChen2014a,*GangChen2014b}. For instance, in the electron CMI on a pyrochlore lattice where the electrons are localized on the tetrahedral clusters, it was shown that~\cite{GangChen2014a}, due to the sub-gap charge fluctuation, the charge sector is in a Coulomb liquid phase with emergent gapless $U(1)$ gauge photon and fractional charge excitation.

\begin{figure}[!h]
\centering
\includegraphics[width=\columnwidth]{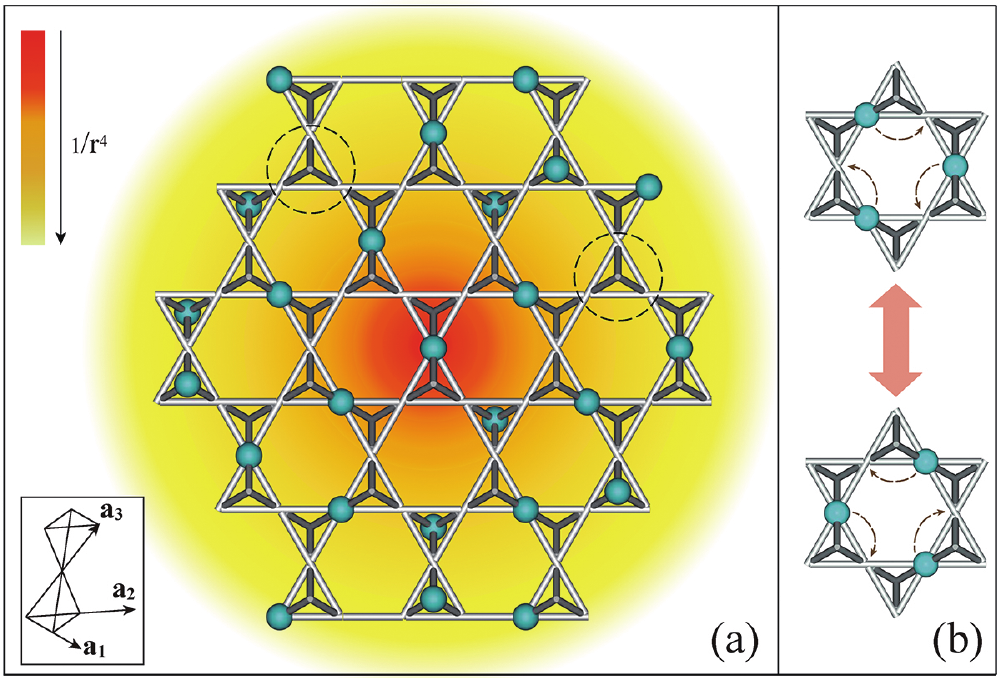}
\caption{(Color online). (a) Snapshot of the boson configuration on the pyrochlore lattice at CMI-1/4, viewed on a Kagome plane. Removing one boson creates two defect tetrahedra (marked with  dashed circles). The intensity of the contour background stands for the $1/r^{4}$ 
decay of the electric field correlation for the Coulomb liquid phase, 
where the electric field component normal to the Kagome plane is chosen. 
The vectors ${\bf a}_{1,2,3}$ are the primitive vectors of the FCC Bravais lattice. (b) The sub-gap boson density fluctuations of the CMI-1/4, happening as the collective tunneling of three bosons around the perimeter of the elementary hexagon.}
\label{fig:schematic}
\end{figure}

The existing proposals for CMIs focus on the electron systems as the real materials are electronic systems. Due to the fermionic statistics, however, the relevant theoretical models often suffer from a fermion sign problem and are prohibited from controlled investigation via quantum Monte Carlo (QMC) simulations. Furthermore, the electron models are  complicated by the presence of the spin carrying excitations. Nevertheless, the hallmark of the CMIs -- particles being localized on the cluster units with sub-gap fluctuations -- is independent of the particle statistics. Therefore, in this paper, we turn our attention to bosons and study the hard-core boson models on a pyrochlore lattice with partial fillings. We show that the bosonic pyrochlore system supports the Coulomb liquid phase with emergent $U(1)$ gauge structure in the bosonic CMI. The emergent and exotic physics in the charge sector of the electronic CMI is fully retained in the corresponding bosonic CMI on the same lattice. 

\paragraph*{Model.}
Our model is defined on the pyrochlore lattice with the Hamiltonian
\begin{equation}
H = \sum_{\langle i,j \rangle}\big[{ -t} (b^{\dagger}_i b^{\phantom\dagger}_j + h.c.) + V n_i n_j \big] - \mu \sum_i  n_i,
\label{eq:model}
\end{equation}
where $b^\dagger$ ($b$) is the hard-core boson creation (annihilation) operator. By definition, we exclude the double occupation of the bosons on a single lattice site. $t$ ($V$) is the nearest neighbor boson hopping (repulsion), and we set $V=1$ as the energy unit throughout the paper.

\begin{figure}[tp]
\centering
\includegraphics[width=\columnwidth]{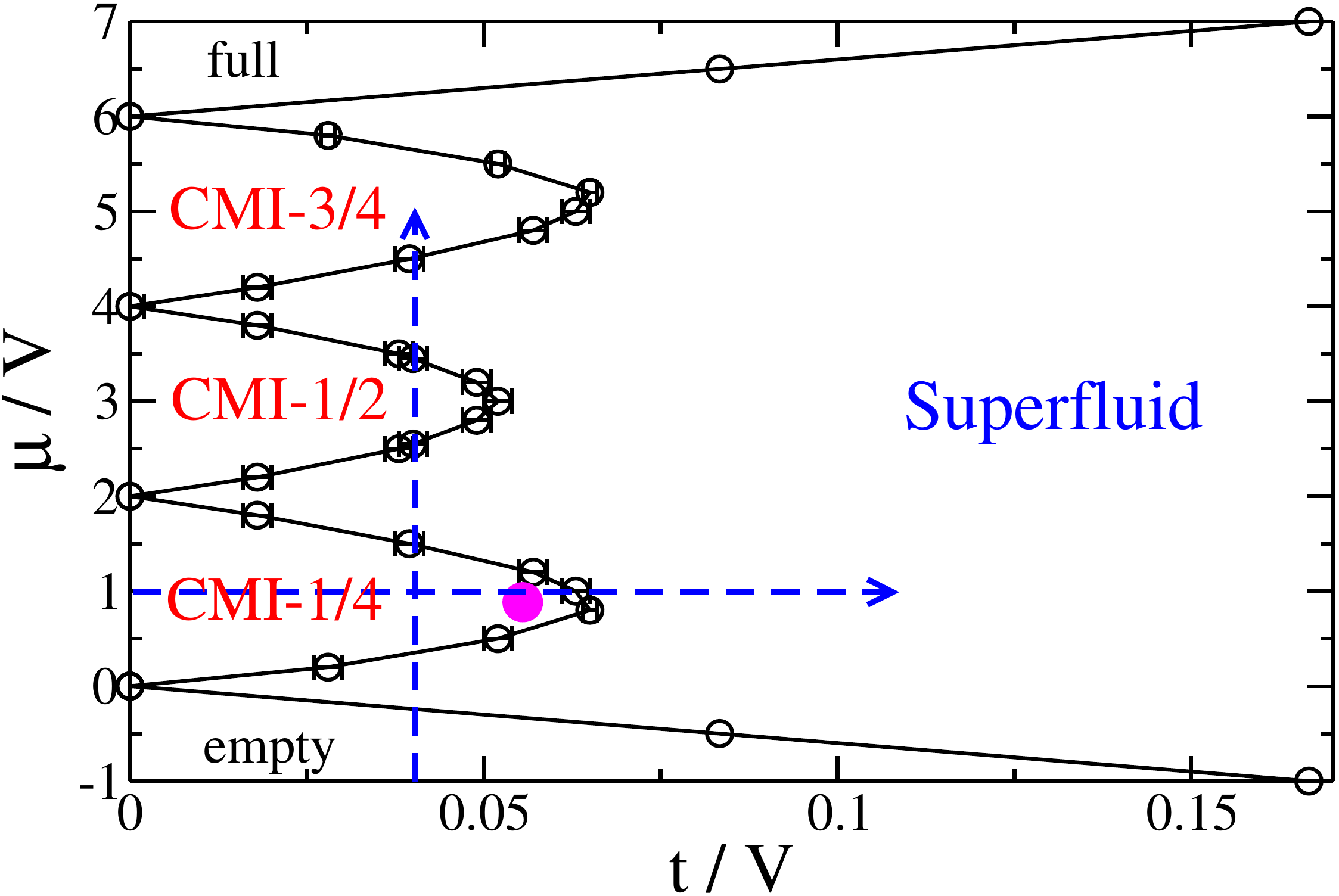}
\caption{(Color online). Phase diagram of the hard-core boson model on a pyrochlore lattice. CMIs are at $1/4$, $1/2$ and $3/4$ boson fillings, respectively. Phase boundaries are determined in QMC simulations with system sizes $4\times L^{3}$. The physical properties along 
the (blue) dashed paths and at the (magenta) solid dot are plotted in Figs.~\ref{fig:fig3}, \ref{fig:electricfieldcorr} and \ref{fig:specificheat}.
 }
\label{fig:phasediagram}
\end{figure}

We first clarify the nomenclatures associated with this model and its ground state in the literature~\cite{Hermele2004,Banerjee2008,Kato2014}. The model in Eq.~\ref{eq:model} was originally proposed as the spin-1/2 XXZ model on a pyrochlore lattice in Ref.~\onlinecite{Hermele2004}, if the hard-core boson variables are transformed into spin operators via $b^{\dagger}_i =  S_i^x + i S_i^y$ and $n_i -1/2 = S_i^z$. The hard-core boson model at 1/2 boson filling gives a Coulomb liquid CMI phase when the inter-site repulsion $V$ is strong enough to localize the bosons in the tetrahedra with 
two bosons per tetrahedron~\cite{Hermele2004,Banerjee2008,Kato2014}. 
This insulating Coulomb liquid phase at 1/2 filling is referred as ``CMI-1/2'' (see Fig.~\ref{fig:phasediagram}) in this letter. 
In the spin language, the ``CMI-1/2'' phase is the well-known quantum spin ice phase whose low-energy property is described by the compact $U(1)$ gauge field coupled with gapped bosonic {\it spinons} that carry quantum nubmer $S^z=1/2$~\cite{Savary12,Sungbin12,Chang2012,Shannon2012a,*Shannon2012b,Huang2014,Gingras&McClarty2014},
and the boson localization condition with two bosons in each tetrahedron is essentially the ``two-up two-down'' spin ice rule~\cite{Hermele2004}.

In this letter, we carry out a large-scale world-line QMC simulation to investigate the full phase diagram of Eq.~\ref{eq:model} in a more general parameter space. We vary both $t/V$ and $\mu/V$ to access different interaction strength as well as the boson fillings other than the $1/2$ filling. We reveal two more CMIs at $1/4$ and $3/4$ fillings (see the schematic plot of the CMI at 1/4 filling in Fig.~\ref{fig:schematic} (a) and the phase diagram in Fig.~\ref{fig:phasediagram}) and verify both of them are in Coulomb liquid phase with emergent compact QED description of the low-energy property. 

\paragraph*{Numerical methods.}
We employ large-scale worm-type QMC technique to simulate the Hamiltonian in Eq.~\ref{eq:model} on the pyrochlore lattice. The simulations are performed in grand canonical ensemble. The worm algorithm is based on the path-integral formulation of the quantum many-body action and works in continuous imaginary time without the Trotter error~\cite{Prokofev1998a,Prokofev1998b,Pollet2012}. In the simulation, we choose the linear lattice size $L$ in the range of $4 \le L \le 16$, larger than those in previous studies~\cite{Banerjee2008,Kato2014}. The inverse temperatures goes as large as $\beta=100L$, while simulations at even lower temperatures were also performed for some cases.

To determine the phase boundary between the superfluid and CMIs, we measure both the superfluid density $\rho_{S}=\langle W^{2}_{\bold{a}_{1}} + W^{2}_{\bold{a}_{2}} + W^{2}_{\bold{a}_{3}} \rangle / (3tL\beta)$ and the boson particle density $\rho=\langle N \rangle/(4L^3)$, where $\langle N \rangle$ is the expectation value of the number of bosons and $W_{\bold{a}_{i}}$, with $i=1,2,3$, are the winding numbers~\cite{Pollock1987} along the three primitive vector directions shown in Fig.~\ref{fig:schematic}a.

\paragraph*{Phase diagram.}
The full phase diagram obtained from our simulations is presented in Fig.~\ref{fig:phasediagram}. When $t/V$ is small, two CMIs at 1/4 and 3/4 boson fillings in addition to the known CMI at 1/2 filling are found. 
Here, the ``CMI-1/4'' (``CMI-3/4'') contains one (three) bosons in each tetrahedral cluster. 
When $t/V$ is large, superfluid phase is obtained. 

\begin{figure}[htp]
\centering
\includegraphics[width=\columnwidth]{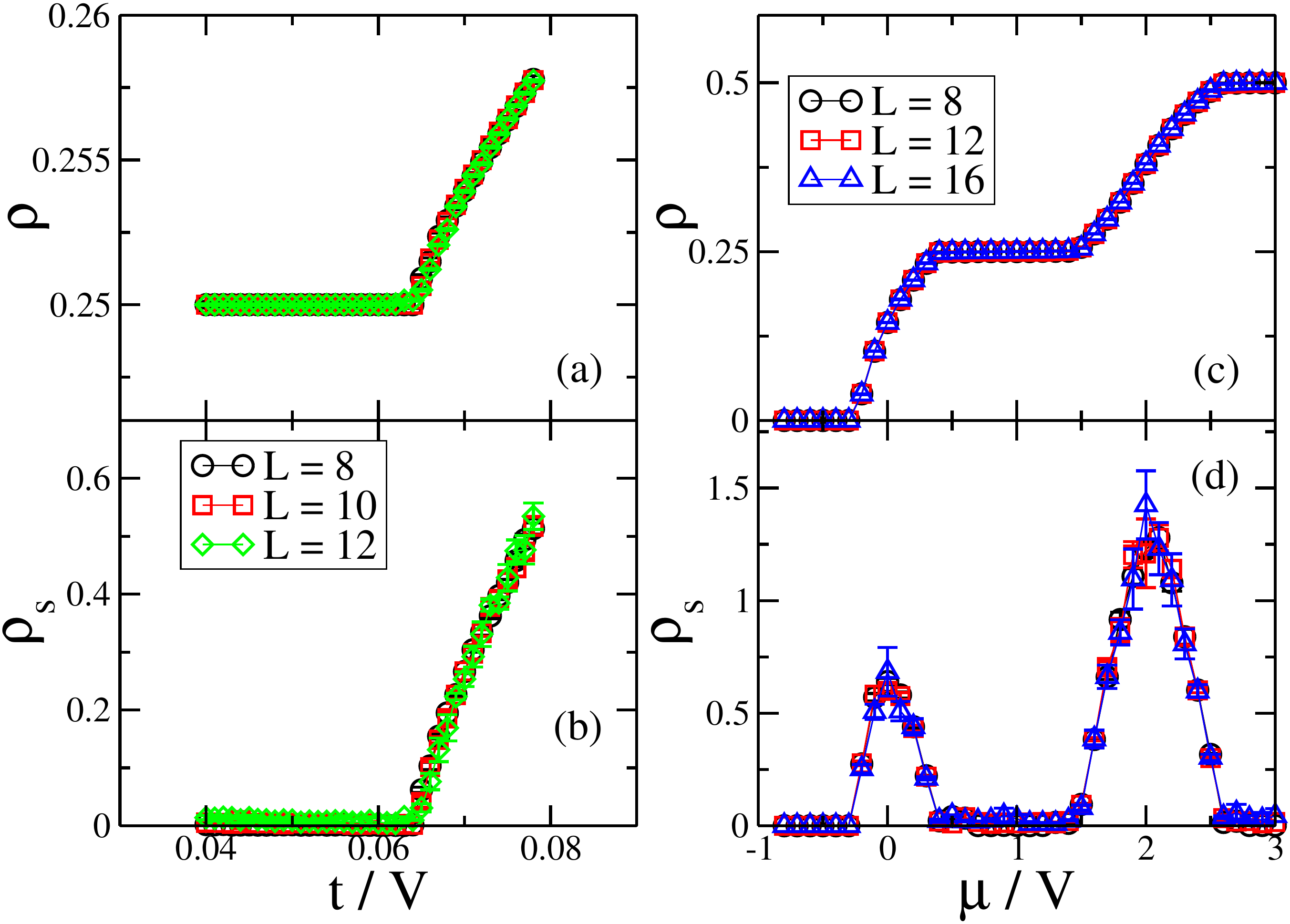}
\caption{(Color online). (a) Boson density $\rho$ and (b) superfluid density $\rho_s$ as a function of interaction $t/V$ along the dashed horizontal path in Fig.~\ref{fig:phasediagram}, with $\mu/V=1$. (c) Boson density $\rho$ and (d) superfluid density $\rho_s$ as a function of chemical potential $\mu/V$ along the dashed vertical path in Fig.~\ref{fig:phasediagram}, with $t/V=0.04$.}
\label{fig:fig3}
\end{figure}

In Figs.~\ref{fig:fig3}a and b, we depict the boson density $\rho$ and the superfluid density $\rho_{S}$ as a function of interaction $t/V$ for a fixed chemical potential $\mu/V=1$, i.e., the system is inside the CMI-1/4 when $t/V$ is small. One sees in the superfluid, $\rho_{S}$ is finite while $\rho$ is strongly fluctuating. The phase boundary of the superfluid is signalled by the vanishing of $\rho_{S}$. In Figs.~\ref{fig:fig3}c and \ref{fig:fig3}d, we depict the boson density $\rho$ and the superfluid density $\rho_{S}$ as a function of chemical potential $\mu/V$ for a fixed interaction strength $t/V=0.04$, i.e., the system is inside the CMI-1/4 and CMI-1/2 when $\mu/V$ is within the Mott regime. In the Mott regime, the boson density is fixed while the superfluid density is zero. A fixed boson density implies that the compressibility $\partial\rho/\partial\mu$ is zero and the system is hence Mott insulating. The Mott insulating phase is terminated at the places where the compressibility is non-zero. In all the cases that we simulated in Fig.~\ref{fig:phasediagram},
and Fig.~\ref{fig:fig3}, the phase boundary of the Mott insulators coincides with the phase boundary of the superfluid. Therefore, there is no other intermediate phase between the superfluid and the Mott insulators.

One can understand the phase diagram in Fig.~\ref{fig:phasediagram} starting from the limit $t\rightarrow 0$. The superfluid extends down to the $t=0$ at even integer values of $\mu/V$, where there is no energy cost to the addition or removal of one bosons onto the tetrahedral unit, 
and hence superfluidity occurs at arbitrarily small hopping $t$. 
Notice that adding (removing) one boson to the system creates two particle-like (hole-like) defect tetrahedra, as shown in Fig.~\ref{fig:schematic}a for the CMI-1/4. 
Here the particle-like (hole-like) defect tetrahedron refers to the tetrahedron that contains one more (less) boson than the boson number demanded by the Mott localization. The Mott lobe at the finite hopping $t$ is understood as the condensation of the mobile particle-like or hole-like defect tetrahedra. As $t$ is increased from zero for a fixed chemical potential inside the Mott regime, 
the kinetic energy gain of the particle-like or hole-like defect tetrahedra increases.
When the kinetic energy gain eventually balances the interaction energy cost, 
the particle-like or hole-like excitations would like to condense, giving rise to superfluidity. Increasing $t$ reduces the chemical potential width of the Mott regime and leads to the lobe-like phase boundary in Fig.~\ref{fig:phasediagram}. 
 
At first looking, our phase diagram and the Mott lobes are quite similar to the ones in the conventional boson Hubbard model~\cite{Fisher89}, but there are key distinctions. In conventional Mott insulators, the boson is at integer fillings and the on-site Hubbard interaction freezes the boson motion and localizes the bosons on the lattice sites. In contrast, the CMIs in Fig.~\ref{fig:phasediagram} are at fractional fillings and it is the inter-site repulsion that localizes them, moreover, the bosons are localized inside the tetrahedral clusters of the pyrochlore lattice.  
For the CMI-1/4 (CMI-3/4) in our phase diagram, one boson (three bosons) is (are) localized on each tetrahedral unit. 
 
More important difference between the conventional Mott insulators and the CMIs in Fig.~\ref{fig:phasediagram} lies in {\it the low-energy density fluctuations}. In conventional Mott insulators, the low-lying density fluctuation that conserves the total boson number is the particle-hole excitation and costs an energy of the order of the Mott gap. In the CMIs, in addition to the similar type of particle-hole excitation as in the conventional Mott insulators, there exist boson density fluctuations far below the Mott gap. As shown in Fig.~\ref{fig:schematic}b, for the CMI-1/4, this sub-gap boson density fluctuation arises from the collective tunnelling of three bosons around the perimeter of the elementary hexagon on the pyrochlore lattice. This collective tunnelling is a third order process of the nearest neighbor hopping $t$ in the Mott regime and costs an energy of $\mathcal{O}(t^3/V^2)$. This process preserves {\it the center of mass} of the three bosons and is hence below the Mott gap. Like in the CMI-1/2
(or quantum spin ice in the spin language) \cite{Hermele2004}, it is the sub-gap density fluctuation that gives rise to the Coulomb liquid phase with emergent compact $U(1)$ QED description at low energies for both the CMI-1/4 and the CMI-3/4~\cite{Bergman06,GangChen2014a}. 

\paragraph*{Coulomb liquid phase of the CMI-1/4.} Now we focus on the CMI-1/4 and provide two  numerical evidences, {\it i.e.} the correlation function of the emergent electric field, 
as well as the temperature dependence of the heat capacity and entropy,
to support the compact $U(1)$ QED nature of the Coulomb liquid phase. 

In the compact QED description of the Coulomb liquid phase \cite{Bergman06,GangChen2014a}, the emergent electric field operator is related to the boson density operator as ${\bf E} ({\bf r}) = \sum_{i \in {\bf r} } ( n_i - 1/4 ) {\bf e}_i$, where the sum is over the four lattice sites on the tetrahedron centered at ${\bf r}$ and ${\bf e}_i$ is the unit vector that points inwards (outwards) from ${\bf r}$ if ${\bf r}$ belongs to the up-pointing (down-pointing) tetrahedron. Using the compact QED action appropriate for the low-energy properties of the Coulomb liquid phase \cite{Hermele2004}, it is readily to evaluate the equal-time electric field correlation and obtain 
\begin{eqnarray}
\langle E_{\mu} ({\bf k}, \tau) E_{\nu} ({\bf k}',\tau) \rangle & = &
( \delta_{\mu\nu} {\bf k}^2 - k_{\mu} k_{\nu})\delta({\bf k} + {\bf k}') 
\nonumber \\
&& \times
\frac{ (2\pi)^4  K_c}{2 v_p |{\bf k}|}\coth (\frac{ \beta v_p | {\bf k} |}{2}),
\label{eq:electricfieldcorrelation}
\end{eqnarray}
where $\mu,\nu = x,y,z$ are the components of the electric field, 
$K_c$ is a proportionality parameter that is determined numerically, 
$v_p$ is the speed of the emergent $U(1)$ gauge photon, 
and $\beta$ is the inverse temperature.  In the zero temperature limit ($\beta \rightarrow \infty$), the above correlation turns into the $1/r^4$ dipolar-like correlation in real space \cite{Hermele2004} (see Fig.~\ref{fig:schematic}a). 

\begin{figure}[htp]
\centering
\includegraphics[width=\columnwidth]{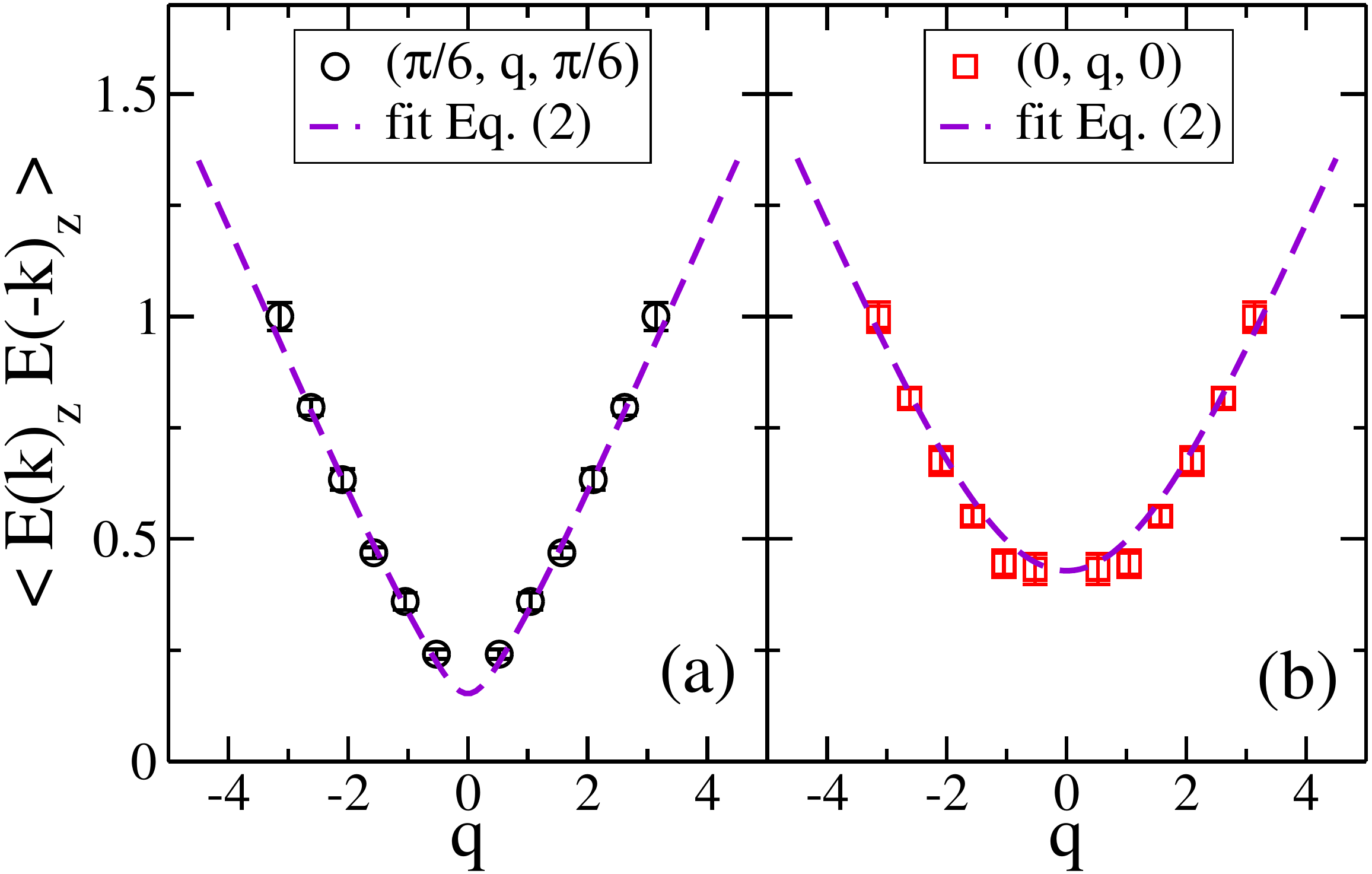}
\caption{(Color online). Electric field correlators at $z$ direction. Dashed curves
are fitted theoretical results using Eq.~\ref{eq:electricfieldcorrelation}. We choose
 $t/V=0.055$, $\mu/V=1$ inside the CMI-1/4
 with system size $L=12$ and inverse temperature $\beta=1000$.} 
\label{fig:electricfieldcorr}
\end{figure}

In Fig.~\ref{fig:electricfieldcorr}, we depict our numerical results for the electric field correlation $\langle E_{z}(\mathbf{k})E_{z}(-\mathbf{k})\rangle$ along different momentum cuts of Brillouin zone. We have chosen the path $\mathbf{k}=(\frac{\pi}{6},q,\frac{\pi}{6})$ in Fig.~\ref{fig:electricfieldcorr}a to fit our data against Eq.~\ref{eq:electricfieldcorrelation} and obtain the photon velocity $v_p = 0.0020(5) a$, where $a$ is the FCC lattice constant. The curve in Fig.~\ref{fig:electricfieldcorr}b is 
plot of Eq.~\ref{eq:electricfieldcorrelation} using the fitted parameters from Fig.~\ref{fig:electricfieldcorr}a, along the path $\mathbf{k}=(0,q,0)$. The fits and data match prefectly in Fig.~\ref{fig:electricfieldcorr}a and b and it provides a unbiased test of the data as only one set of parameters is used to fit the data along different momentum cuts. 

\begin{figure}[h]
\centering
\includegraphics[width=\columnwidth]{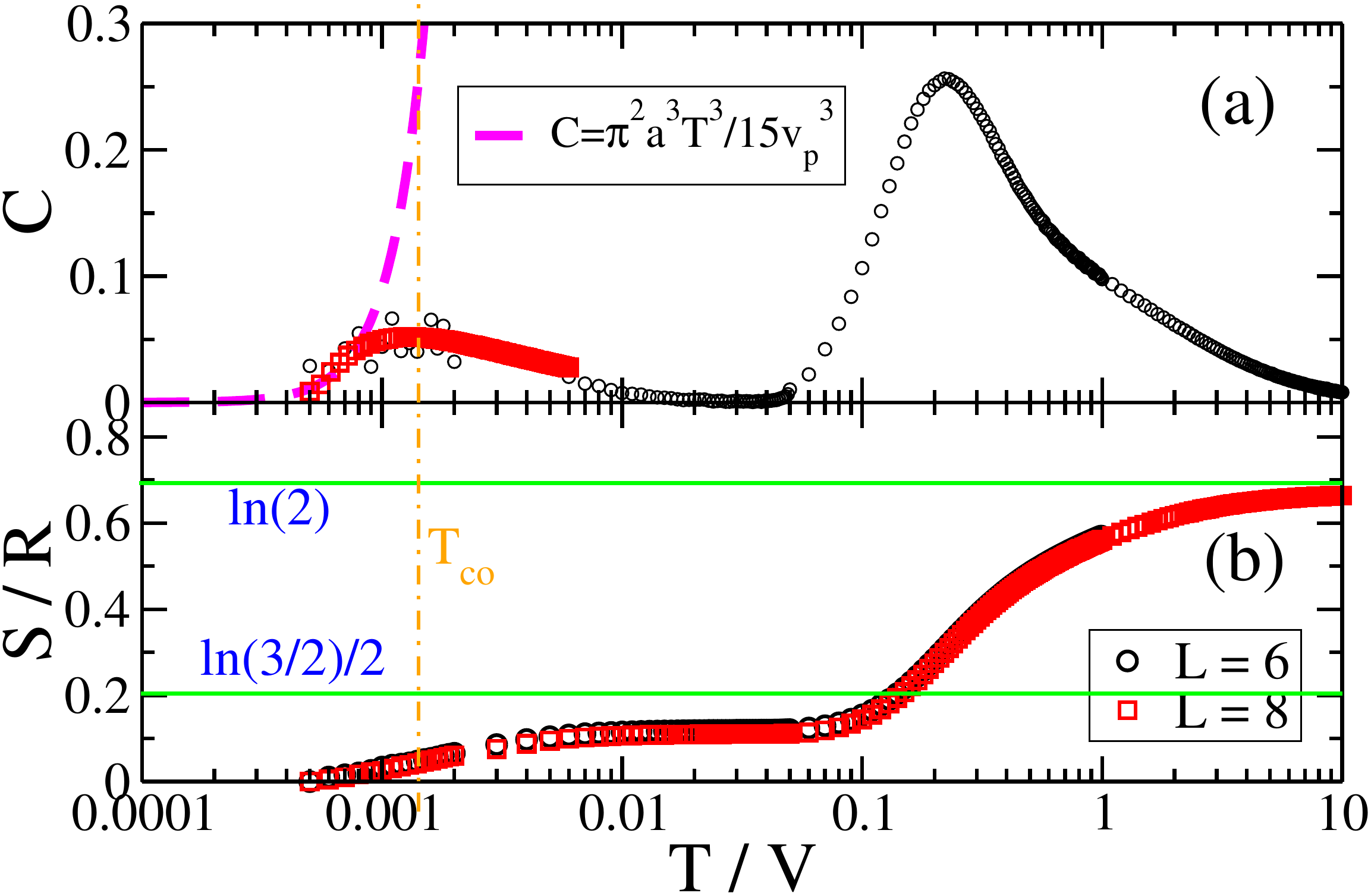}
\caption{(Color online). Temperature dependence of (a) the specific heat $C$ and (b) the entropy $S$ for $t/V=0.055$, $\mu/V=1$ inside the CMI-1/4. The (magenta) dashed lines in (a) is the fit to specific heat of Coulomb liquid phase. The (orange) dotted-dash lines highlight the crossover temperature $T_{\text{co}}$ \cite{Note}. The (green) horizontal lines denotes the Pauling entropy of classical spin ice $S_{p}={\text R}\frac{1}{2}\ln(\frac{3}{2})$, 
and the entropy of free spins $S={\text R}\ln(2)$ at the high temperature limit.}
\label{fig:specificheat}
\end{figure}

Due to the gapless gauge photon with dispersion $\epsilon_{\bf k} = v_p |{\bf k}|$ in the Coulomb liquid phase, we expect that the heat capacity $C(T) \approx T^3  {\pi^2 a^3}/{(15 v_p^3)}$ for the CMI-1/4 at zero temperature limit, where $a$ is the FCC lattice constant and $v_p$ is the photon speed. As shown in Fig.~\ref{fig:specificheat}a, this $T^3$ dependence of the low-temperature heat capacity is clearly obtained for the CMI-1/4 for $T < 0.002V$, and we fit to the theoretical prediction and obtain $v_p = 0.0017(4) a$. As there is no continuous symmetry breaking and thus no gapless Goldstone mode in the CMI-1/4, the $T^3$ heat capacity is a strong indication of the presence of linearly dispersive gapless gauge photon~\footnote{Similar temperature dependence is also observed for heat capacity in the CMI-1/2 and is  
consistent with Ref.~\onlinecite{Kato2014}}. Moreover, the photon speed extracted from heat capacity is consistent with the one obtained previously from electric field correlation. These can be served as two independent evidences that the CMI-1/4 is indeed well described by the compact $U(1)$ QED. In Fig.~\ref{fig:specificheat}b, we depict the finite temperature behavior of the entropy for the CMI-1/4 phase \cite{Note}.

\paragraph*{Discussion.} 
All exisiting theoretical and numerical works except Ref.~\onlinecite{Bergman06} consider the CMI-1/2 state of the hard-core boson model (or quantum spin ice in the spin language) on the pyrochlore lattice.  
We have carried out the first numerical study of the full phase diagram of the model
and provide the numerical evidence for the bosonic CMIs in the phase diagram. 
Our numerical results of Coulomb liquid phases of the bosonic CMIs are relevant to 
the charge sector physics of electron CMIs in the pyrochlore materials \cite{GangChen2014a,Abd-Elmeguid2004}. One future direction is to include disorders and to understand if any intermediate Bose-glass-like phase occurs between the CMIs and the superfluid. If such a phase do exist for a disordered system, the physical property might be very different from the Bose glass for the conventional boson Hubbard model with disorders \cite{Fisher89}.

\paragraph*{Acknowledgements.-} 
We thanks Leon Balents and Roderich Moessner for the encouragement and positive comments on our results and manuscript. We acknowledge Xiao Yan Xu for technical help in preparing Fig.~\ref{fig:schematic}.  
This work was supported by the Natural Science Foundation of China (NSFC) under Grant Nos. 11147013 and 11405003 (JPL), the Center for International Collaboration at Institute of Physics, Chinese Academy of Sciences (GC), the NSFC under Grant No. 11275185 and the National Key Basic Research Support Foundation of China (NKBRSFC) under Grant No. 2011CB921300 (YJD), and the National Thousand-Young-Talents Program of China (ZYM). Most of the simulations were carried out at National Supercomputer Center in Tianjin on the platform TianHe-1A.

\bibliography{draft1}

\end{document}